\documentclass[a4paper,11pt]{article}
\pdfoutput=1 

\usepackage{pub} 

\usepackage[T1]{fontenc} 

\makeatletter
\def\@fpheader{\relax}
\makeatother

\title{\boldmath Preheating in radiative corrections to $\phi^4$ inflation with non-minimal coupling in Palatini formulation}

\author[]{Nilay Bostan}

\affiliation{Department of Physics and Astronomy, University of Iowa, \\52242, Iowa City, IA, USA }

\emailAdd{nilay-bostan@uiowa.edu}

\abstract{We discuss the impact of the preheating stage in radiative corrections due to interaction of the inflaton to fermions to $\phi^4$ inflation with non-minimal coupling in Palatini formulation. In Palatini inflation with large non-minimal coupling the field is allow to return to the plateau region during the reheating stage, so the average equation of state per oscillations is closer to $-1$ than to $1/3$. The incursion in the plateau leads, however, to a highly efficient tachyonic instability able to reheat the Universe in less than one e-fold. By taking into account prescription II discussed in the literature, in the wide range of $\kappa-\xi$, we figure out spectral index $n_s$ and tensor-to-scalar ratio $r$ which are compatible with the data given by the Keck Array/BICEP2 and Planck collaborations. }


\begin{document}
\maketitle
\flushbottom

\section{Introduction}
Inflation \cite{Guth:1980zm,Linde:1981mu,Albrecht:1982wi,Linde:1983gd} is an early period of nearly exponential expansion of the universe and it has become a solution for the several shortcomings such as horizon, flatness and unobserved magnetic monopoles until around 1980 when the inflationary theory proposed. The theory of cosmic inflation gives an acceptable explanation of the large scale homogeneity of the universe as well as primordial density perturbations that grow into cosmic structure. These primordial perturbations evolve to produce the observed large scale structure and cosmic microwave background (CMB) temperature anisotropy. In addition to this, several inflationary models have been suggested \cite{Martin:2013tda} and most of them defining by the slow-rolling scalar field $\phi$, called the inflaton. Predictions of these models are currently being tested by the polarization observations and CMBR temperature anisotropies \cite{Aghanim:2018eyx,Akrami:2018odb}. In particular, the last results release of the Keck Array/BICEP2 and Planck collaborations \cite{Ade:2018gkx} cast constraints robustly on the tensor-to-scalar ratio $r$,  which provides an explanation of the amplitude of primordial gravitational waves and to the scale of inflation. As a result, the predictions of the models of simple monomial inflation are ruled out at $2 \sigma$ level, so non-minimally coupled to gravity models become as the most popular ones. 

In this work, we take into account that the models of inflation with a non-minimal coupling to gravity ($\xi \phi^2 R$), where $\xi$ is the non-minimal coupling parameter, $\phi$ is the scalar field (inflaton) and $R$ is the Ricci scalar, $\xi \phi^2 R$ term is essential to provide the renormalizability of the scalar field theory in curved space-time \cite{Callan:1970ze,Freedman:1974ze,Buchbinder:1992rb} and the predictions of the inflationary models are changed significantly according to the coefficient of this coupling term \cite{Martin:2013tda}. We show that the presence of the non-minimal coupling parameter $\xi$, how the values of $n_s$ and $r$ change for preheating in radiative corrections due to interaction of the inflaton to fermions to $\phi^4$ inflation with non-minimal coupling in Palatini formulation for prescription II. In literature, a large number of articles studied to the inflation with non-minimal coupling in Metric formalism \cite{Bezrukov:2010jz,Bostan:2018evz,Bezrukov:2007ep}. In particular, the scenario where the Standard Model Higgs scalar \cite{Bezrukov:2007ep} is the inflaton which is the most favorite one. Furthermore, in Metric formulation, all models asymptote to a universal attractor \cite{Kallosh:2013tua}, called Starobinsky model for the large values of the $\xi$, independence of the original scalar potential. On the other hand, the attractor behaviour of the Starobinsky model is lost in the Palatini formulation and $r$ can be smaller compared with the Metric one for the Palatini formulation \cite{Bauer:2008zj,Jarv:2017azx}. Also, consideration for the gravitational degrees of freedom is necessary in the presence of non-minimal couplings to gravity. In the metric formulation of gravity the independent variables are the metric and its first derivatives \cite{Padmanabhan:2004fq,Paranjape:2006ca}, while in the Palatini formulation the independent variables are the connection and the metric \cite{Attilio, Einstein, Ferraris}. The predictions of the two formalisms correspond to the same equations of motion, so they describe equivalent physical theories. However, in case of non-minimal couplings between gravity and matter, such equivalence is disappear and the two formulations illustrate different gravity theories \cite{Bauer:2008zj,York:1972sj,Tenkanen:2017jih,Rasanen:2017ivk,Racioppi:2017spw,Tamanini:2010uq, Racioppi:2019jsp}. In the literature, Palatini formulation of inflation with non-minimal coupling debated in refs. \cite{Bauer:2008zj,Rasanen:2017ivk,Racioppi:2017spw,Fu:2017iqg,Gumjudpai:2016ioy,Markkanen:2017tun,Rubio:2019ypq}. Palatini self-interaction potential $V(\phi)$ analyzed in ref. \cite{Bauer:2008zj} and they figured out observational parameters $n_s\simeq0.968$ and $r\simeq 10^{-14}$ in the large field limit and also, Palatini Higgs inflation calculated in ref. \cite{Rasanen:2017ivk} and they found the range of tensor-to-scalar ratio $1\times10^{-13}<r<2\times 10^{-5}$, therefore in Palatini formulation $r$ takes too tiny values. In addition to this, it was showed that the radiative corrections to inflationary potential might play pivotal role \cite{Marzola:2015xbh,Marzola:2016xgb,Dimopoulos:2017xox}, in case of non-minimal couplings to gravity, generating the
Planck scale dynamically \cite{Kannike:2015kda}.

In this paper, we study the impact of the preheating stage in Palatini radiatively corrected $\phi^4$ inflation for prescription II and inflaton to fermions coupling. As compared to the metric formulation, the entropy production in Palatini Higgs inflation appears significantly more effective \cite{Rubio:2019ypq}, decreasing the number of e-folds required to solve the flatness and horizon problems and producing to a less spectral tilt for primordial density perturbations. Ref. \cite{Rubio:2019ypq} showed that after the inflation, the slow decay of the Higgs oscillations lets the field return to the plateau of the potential periodically during the reheating stage, in the large field limit, the effective mass of Higgs occurs negative, letting for the exponential creation of Higgs excitations. As a consequence, the prehating stage of the Palatini Higgs inflation is primarily instantaneous and also this case decreases the number of $N_*$ required to solve the hot big bang shortcomings \cite{Rubio:2019ypq}. The paper is organized as follows: non-minimal inflation with Palatini formalism is presented in section \ref{inf}. In section \ref{rad}, we explain the radiative corrections to the potential for the renormalization prescription II for the fermions coupling. In section \ref{pres}, we display the impact of the preheating stage in Palatini radiatively corrected $\phi^4$ inflation for prescription II and inflaton to fermions coupling numerically and finally, we discuss our results in section \ref{conc}. 

\section{Non-minimal inflation in Palatini formulation} \label{inf}
Assuming the following Lagrangian density for a scalar-tensor theory in the Jordan frame with non-minimally coupled scalar field $\phi$:
\begin{equation} \label{vjphi}
\frac{\mathcal{L}_J}{\sqrt{-g}}=\frac12F(\phi)R-\frac12g^{\mu\nu}\partial_{\mu}\phi\partial_{\nu}\phi-V_J(\phi)\,,
\end{equation}
where the subscript $J$ indicates that the Lagrangian is described in a Jordan frame and $F(\phi)=1+\xi \phi^2$ with a canonical kinetic term and a potential $V_J(\phi)$. We consider the units where the reduced Planck scale $m_P=1/\sqrt{8\pi G}\approx2.4\times10^{18}\text{ GeV}$ is fixed equal to unity, thus we require $F(\phi)\to1$ after inflation and to avoid repulsive gravity, we suppose $F(\phi)>0$. This property of $F(\phi)$ is independent on the formulations of gravity for example Metric and Palatini.

In the metric formulation the connection is described as a function of metric tensor called as Levi-Civita connection  ${\bar{\varGamma}={\bar{\varGamma}}(g^{\mu \nu})}$:
\begin{equation} \label{vargammametric}
\bar{\varGamma}_{\alpha \beta}^{\lambda}=\frac{1}{2}g^{\lambda \rho} (\partial_{\alpha}g_{\beta \rho}+\partial_{\beta}g_{\rho \alpha}-\partial_{\rho}g_{\alpha \beta}).
\end{equation}
On the contrary, $g_{\mu \nu}$ and $\varGamma$ are independent variables  in the Palatini formalism, and the unique constraint is that the connection is torsion-free, $\varGamma_{\alpha \beta}^{\lambda}=\varGamma_{\beta \alpha }^{\lambda}$. By
solving the EoM, we obtain \cite{Bauer:2008zj}
\begin{eqnarray}\label{vargammapalatini}
\Gamma^{\lambda}_{\alpha \beta} = \overline{\Gamma}^{\lambda}_{\alpha \beta}
+ \delta^{\lambda}_{\alpha} \partial_{\beta} \omega(\phi) +
\delta^{\lambda}_{\beta} \partial_{\alpha} \omega(\phi)- g_{\alpha \beta} \partial^{\lambda} \omega(\phi),
\end{eqnarray}
where 
\begin{eqnarray}
\label{omega}
\omega\left(\phi\right)=\ln\sqrt{F(\phi)}.
\end{eqnarray}
Due to the fact that the connections (eqs. \eqref{vargammametric} and \eqref{vargammapalatini}) are different, the metric and Palatini fomalisms correspond to two different theories of gravity. On the one hand, we can explain the differences by taking account the problem in the Einstein frame by means of the conformal transformation.

In order to figure out the observational parameters, it is more efficient way to switch the Einstein frame by using a Weyl rescaling $g_{E, \mu \nu}=g_{J, \mu \nu}/F(\phi)$, the Einstein frame Lagrangian density becomes \cite{Fujii:2003pa}

\begin{equation} \label{LE}
\frac{\mathcal{L}_E}{\sqrt{-{g_E}}}=\frac12{R_E}-\frac{1}{2Z(\phi)}g_E^{\mu\nu}\partial_{\mu}\phi\partial_{\nu}\phi-V_E(\phi)\,,
\end{equation}
where
\begin{equation} \label{Zphi}
Z^{-1}(\phi)=\frac{1}{F(\phi)}\,,\qquad
V_E(\phi)=\frac{V_J(\phi)}{F(\phi)^2}\,,
\end{equation}
in the Palatini formalism. By making a field redefinition
\begin{equation}\label{redefine}
\mathrm{d}\sigma=\frac{\mathrm{d}\phi}{\sqrt{Z(\phi)}}\,,
\end{equation}
we find the Lagrangian density for a minimally coupled scalar field $\sigma$ with a canonical kinetic term. As a consequence, for the Palatini formalism, the field redefinition is induced just by the rescaling of the inflaton kinetic term and also it does not include of Jordan frame Ricci scalar. On the other hand, in Metric formalism, the field redefinition consists of the transformation of the Jordan frame Ricci scalar and the rescaling of the Jordan frame scalar field kinetic term \cite{Bauer:2008zj}. Therefore, we can say that the difference between metric and Palatini formalisms correspond to the different definition of $\sigma$ with the different non-minimal kinetic term including $\phi$.

In the large-field limit, for $F(\phi)=1+\xi\phi^2$,  
($|\xi|\phi^2\gg1$), we figure out
\begin{equation} \label{strong}
\phi\simeq\frac{1}{\sqrt{\xi}}\sinh \left(\sigma\sqrt{\xi}\right),
\end{equation}
in the Palatini formalism. By using eq. \eqref{strong}, inflationary potential can be described in terms of $\sigma$, so we can obtain slow-roll parameters in the Palatini formalism for the $|\xi|\phi^2\gg1$ limit in terms of $\sigma$.

Supposing slow-roll, observational parameters for the inflationary dynamics are defined by the slow-roll parameters \cite{Lyth:2009zz},
\begin{equation}\label{slowroll1}
\epsilon =\frac{1}{2}\left( \frac{V_{\sigma} }{V}\right) ^{2}\,, \quad
\eta = \frac{V_{\sigma \sigma} }{V},
\end{equation}
where $\sigma$'s in the subscript indicate derivatives. Observational parameters i.e. the spectral index $n_s$ and the tensor-to-scalar ratio $r$ are expressed in terms of the slow-roll parameters 
\begin{eqnarray}\label{nsralpha1}
n_s = 1 - 6 \epsilon + 2 \eta \,,\quad
r = 16 \epsilon.
\end{eqnarray}
The number of e-folds, in the slow-roll approximation
\begin{equation} \label{efold1}
N_*=\int^{\sigma_*}_{\sigma_e}\frac{V\rm{d}\sigma}{V_{\sigma}}\,, \end{equation}
where the subscript ``$_*$'' indicates quantities when the scale
corresponding to $k_*$ exited the horizon, and $\sigma_e$ is the inflaton
value at the end of inflation, which we define via $\epsilon(\sigma_e) =
1$. 

The amplitude of the curvature power spectrum is given the form
\begin{equation} \label{perturb1}
\Delta_\mathcal{R}=\frac{1}{2\sqrt{3}\pi}\frac{V^{3/2}}{|V_{\sigma}|}.
\end{equation}
The best fit value for the pivot scale $k_* = 0.002$ Mpc$^{-1}$ is $\Delta_\mathcal{R}^2\approx   2.4\times10^{-9}$ \cite{Aghanim:2018eyx} from the Planck results. 

Furthermore, we reproduce slow-roll parameters in terms of the original scalar field $\phi$ for using in the numerical calculations. By using with together eqs. \eqref{redefine} and \eqref{slowroll1}, slow-roll parameters can be figured out in terms of $\phi$ \cite{Linde:2011nh}
\begin{eqnarray}\label{slowroll2}  
\epsilon=Z\epsilon_{\phi}\,, \ \
\eta=Z\eta_{\phi}+{\rm sgn}(V')Z'\sqrt{\frac{\epsilon_{\phi}}{2}},
\end{eqnarray}
where we defined 
\begin{equation}
\epsilon_{\phi} =\frac{1}{2}\left( \frac{V^{\prime} }{V}\right) ^{2}\,, \quad
\eta_{\phi} = \frac{V^{\prime \prime} }{V} .
\end{equation}
In similar, eqs. \eqref{efold1} and \eqref{perturb1} can be found by in terms of $\phi$
\begin{eqnarray}\label{perturb2}
N_*&=&\rm{sgn}(V')\int^{\phi_*}_{\phi_e}\frac{\mathrm{d}\phi}{Z(\phi)\sqrt{2\epsilon_{\phi}}}\,,\\
\label{efold2} \Delta_\mathcal{R}&=&\frac{1}{2\sqrt{3}\pi}\frac{V^{3/2}}{\sqrt{Z}|V^{\prime}|}\,.
\end{eqnarray}
These observable parameters depend on the number of e-folds of inflation required to solve such problems i.e. the flatness and horizon. Following the standard method, we need 

\begin{equation} \label{eq:efold_formula}
1=a_0=\frac{a_0}{a_\mathrm{RH}}\frac{a_\mathrm{RH}}{a_\mathrm{e}}\frac{a_\mathrm{e}}{a_*}a_*
=	\left(\frac{g_{*s\,\mathrm{RH}}}{g_{*s\,\mathrm{now}}}\right)^{1/3} \frac{T_\mathrm{RH}}{T_0}
\frac{k_*}{H_*}\exp\left(\Delta N+N_*\right),
\end{equation}
``$0$'' denotes to the value of the corresponding quantity at the present time and (``RH'') indicates at the end of the reheating stage. The quantity $\Delta N$ indicates the number of e-folds of reheating, $g_{*s}$ is the effective number of entropy degrees of freedom with $g_{*s\,\mathrm{RH}}=g_{*\, \mathrm{RH}}$ and $g_{*s\,\mathrm{now}}=3.94$ \cite{Husdal:2016haj} and $T_0 \simeq 2.7$ K. $T_{RH}$ is the reheating temperature and in ref. \cite{Rubio:2019ypq}, $N_{*}$ is defined in preheating Palatini Higgs inflation which is necessarily instantaneous and after the inflation, almost whole of the background energy density converted to the radiation and by solving eq. \eqref{eq:efold_formula} for the condition of $\Delta N\ll 1$, $N_*$ can be found in that form \cite{Rubio:2019ypq}
\begin{equation} \label{eq2} 
N_*\simeq 54.9- \frac{1}{4} \log \xi,
\end{equation} 
this result is precise to an integer order of $N_*$. In section \ref{pres}, we figure out numerically the impact of the preheating stage in Palatini radiatively corrected $\phi^4$ inflation for prescription II and inflaton to fermions coupling by using eq. \eqref{eq2}.
				
\section{Radiatively corrected $\phi^4$ potential for coupling to fermions in prescription II } \label{rad}
For the description of couplings of the inflaton with other fields, it is necessary for effective reheating, produce to radiative
corrections in the inflationary potential. These corrections can be defined at leading order in that form \cite{Coleman:1973jx,Enqvist:2013eua,Weinberg:1973am}
\begin{eqnarray} \label{CW}
\Delta V(\phi)=\sum\limits_{i}\frac{(-1)^\nu}{64\pi^2}M_i(\phi)^4 \ln\Big(\frac{M_i(\phi)^2}{\mu^2}\Big).
\end{eqnarray}
Here, $\nu$ is $-1$ for fermions, $\mu$ is a renormalization scale and $M_i(\phi)$ correspond field dependent mass. 
				
We consider the inflationary potential for a minimally coupled $\phi^4$ potential interacts to other scalar $\chi$ and to a Dirac fermion $\Psi$ in that form
\begin{eqnarray}\label{lag}
V(\phi,\chi,\Psi)= \frac{\lambda}{4}\phi^4+h\phi\bar{\Psi}\Psi+m_\Psi\bar{\Psi}\Psi+\frac{1}{2}g^2\phi^2\chi^2+\frac{1}{2}m_\chi^2\chi^2 .
\end{eqnarray}
We assume that these approximations
\begin{eqnarray} \label{assume}
g^2\phi^2\gg m_\chi^2, \qquad g^2\gg\lambda, \nonumber\\ \qquad h\phi\gg m_\Psi, \qquad h^2\gg\lambda,
\end{eqnarray} 
the inflationary potential consisting of the Coleman-Weinberg one-loop corrections given by eq. \eqref{CW} can be figured out for the fermions in that form 
\begin{eqnarray}\label{coupling}
				V(\phi)\simeq\frac{\lambda}{4}\phi^4-\kappa\phi^4\ln\left(\frac{\phi}{\mu}\right),
				\end{eqnarray} 
				we can describe the coupling parameter as follows 
				\begin{eqnarray}\label{kappa}
				\kappa\equiv\frac{1}{32\pi^2}\Big|(g^4-4h^4)\Big|.
				\end{eqnarray}
				Here, the potential in eq. \eqref{coupling} is just approximation of the one-loop RG improved effective actions \cite{Okada:2010jf}. 
				
				As discussed in the literature, one of two different prescriptions is the prescription II that is typically using for the computation of radiative corrections \cite{DeSimone:2008ei,Barvinsky:2009fy,Barvinsky:2009ii}. 
				In prescription II, the field dependent masses in the one-loop Coleman-Weinberg potential are described in the Jordan frame, therefore eq. \eqref{coupling} corresponds to the one-loop Coleman-Weinberg potential in the Jordan frame. As a consequence, the Einstein frame potential for interactions of inflaton and fermions in prescription II is described by
				\begin{eqnarray} \label{p2}
				V(\phi)=\frac{\frac{\lambda}{4}\phi^4-\kappa \phi^4 \ln\Big(\frac{\phi}{\mu}\Big)}{(1+\xi\phi^2)^2} .
				\end{eqnarray} 
				We can say that the variation of the value of renormalization scale does not affect the form of potential in eq. \eqref{p2}. The form of the potential only changes with shifting of $\lambda$. As a result, observational parameters do not affect the change of $\mu$ as well. 
				
				\begin{figure}[]
					\centering
					\includegraphics[angle=0, width=15cm]{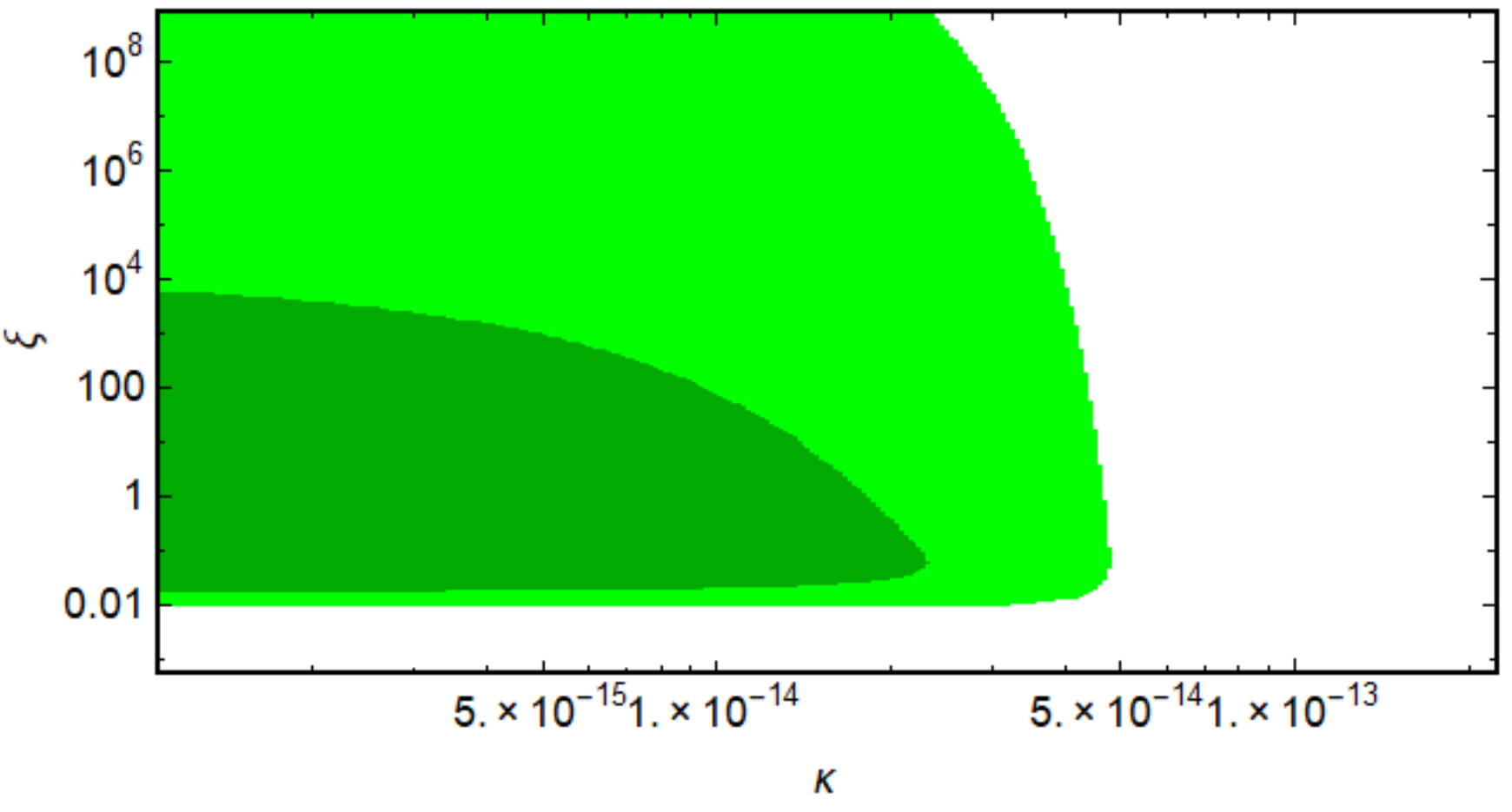}
					
					\
					
					\includegraphics[angle=0, width=17cm]{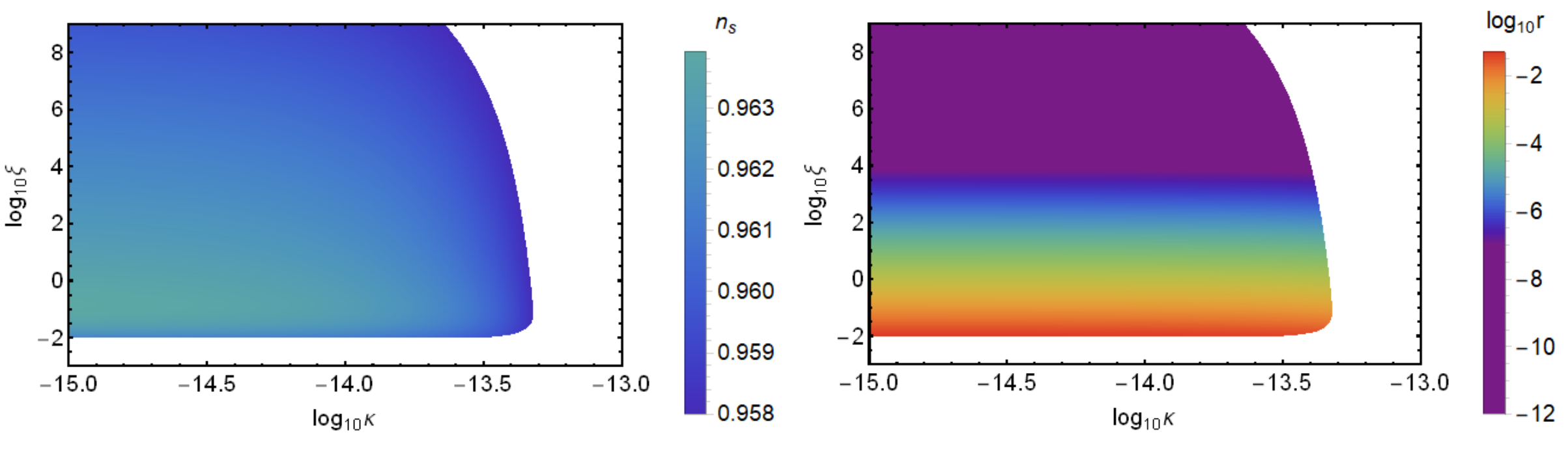}
					\caption{The top figure illustrates in light green (green) the regions in the $\kappa-\xi$ plane for the $n_s$ and $r$ values are inside the $95\%  (68\%) $ CL contours based on data given by the Keck Array/BICEP2 and Planck collaborations \cite{Ade:2018gkx}. Bottom figures display $n_s$ and $r$ values in these regions.}
					\label{fig2}
				\end{figure}

					\section{Inflationary results} \label{pres}
					In this section, we investigate numerically the affect of the preheating stage in Palatini radiatively corrected $\phi^4$ inflation for prescription II and inflaton to fermions coupling on the $n_s$ and $r$ as a function of the coupling parameter $\kappa$ and the non-minimal coupling parameter $\xi$. Figure 1 displays that the region in the coupling parameter and non-minimal coupling parameter plane where the values of $n_s$ and $r$ are agreement with the current measurements. As it can be seen from fig.1, for the values of $10^{-2}\lesssim \xi \lesssim 10^{4}$ and $10^{-15}\lesssim \kappa \lesssim 2.2\times 10^{-14}$, observational parameters can be within the $ 68\% $ CL contour based on data given by the Keck Array/BICEP2 and Planck collaborations and their values are $n_s\simeq 0.963$ and $10^{-7}\lesssim r \lesssim 10^{-2}$. On the other hand, as $\kappa$ increases, it has maximum value for each $\xi$ value and therefore in case of $\kappa>\kappa_{\mathrm{max}}$ values, there has not any solution for providing to the inflationary dynamics. Furthermore, in the range between $10^{4}\lesssim \xi \lesssim 10^{8}$ and $10^{-15}\lesssim \kappa \lesssim 5\times 10^{-14}$, we find $0.958 \lesssim n_s\lesssim 0.961$ and $10^{-12}\lesssim r \lesssim 10^{-7}$, these values are in the $ 95\% $ CL contour based on data given by the Keck Array/BICEP2 and Planck collaborations. As a result, for $10^{4}\lesssim \xi \lesssim 10^{8}$ and $N_*\simeq 52$, we obtain $0.958  \lesssim n_s\lesssim 0.961$. Even though still in $2 \sigma$ confidence limits, these $n_s$ values are slightly disagreed with the observational results given by the Keck Array/BICEP2 and Planck collaborations as well as the values of $r$ is extremely tiny in the large $\xi$ limits. Ref. \cite{Rubio:2019ypq} also found that for the preheating stage of Higgs inflation in Palatini formulation, $n_s\simeq 0.961$ and $r$ values are very tiny in the large $\xi$ values for the $N_*\simeq 51$ and finally, the behaviour of Starobinsky attractor in metric formulation for large $\xi$ values is lost for the potential we take into account. 
					
					\section{Conclusion}\label{conc}
In this paper, we described non-minimal inflation in Palatini formulation in section \ref{inf} and then we briefly presented the radiative corrections to the potential for the renormalization prescription II for couplings to fermion in section \ref{rad}. We investigated the impact of the preheating stage on the observational parameters for this type of potential numerically in section \ref{pres}. 

In general, we found that $r$ values are too small in the large field limit and the behaviour of Starobinsky attractor in metric formulation for the large $\xi$ values is disappear for the potential we considered.  Furthermore, we found that for the cases of $\kappa>\kappa_{\mathrm{max}}$, there has not any solution for providing inflationary dynamics and for the values of $10^{4}\lesssim \xi \lesssim 10^{8}$ and $N_*\simeq52$, $n_s$ values are in $2\sigma$ CL but marginally incompatible with the observational results. 
						
In the large field limit, for the Palatini formulation, the process of entropy production emerges very efficient and leads to the complete reduction of the inflaton condensate in smaller than one e-fold of expansion \cite{Rubio:2019ypq}. As a consequence, in the preheating stage Palatini inflation with the radiative corrections for the fermions coupling in prescription II is necessarily instantaneous and after the end of inflation, almost all of the background energy density converted to the radiation. This decreases the $N_*$ required to solve the common hot big bang shortcomings while outstanding an insignificantly smaller value for the spectral tilt. 
						
Finally, by consideration of $\mathcal{O}(10^{-3})$ accuracy of future precision measurements \cite{Remazeilles:2017szm}, the predictions of Palatini formulation could be distinguished from the metric one within a forthcoming results and supposing a larger values of $r$ is acquired, Palatini formulation can rule out.
\section{Acknowledgements}
The author very thanks Javier Rubio for useful discussions.

\end{document}